\documentclass[%
 reprint,
 amsmath,amssymb,
 aps,
 prl,
]{revtex4-2}
\usepackage{graphicx}% Include figure files
\usepackage{dcolumn}% Align table columns on decimal point
\usepackage{bm}% bold math

\usepackage[colorlinks=true,allcolors=blue]{hyperref}
\newcommand*{\nepg}{$^{21}$Ne(p,$\gamma$)$^{22}$Na }

\usepackage{booktabs}
\usepackage{multirow}
\begin{document}

%\preprint{APS/123-QED}

\title{Improved direct measurement of low-energy resonances in the $^{21}$Ne(p,$\gamma$)$^{22}$Na reaction}

\author{R.\,S.\,Sidhu$^{1,2}$}
\author{F.~Casaburo$^{3,4}$}
\author{E.~Masha$^{5}$}
\author{M.~Aliotta$^{1}$}
\author{C.\,Ananna$^{6,7}$}
\author{L.\,Barbieri$^1$}
\author{F.~Barile$^{8,9}$}
\author{C.\,Baron$^{10,11}$}
\author{D.~Bemmerer$^5$}
\author{A.~Best$^{6,7}$}
\author{R.\,Biassisi$^{10,11}$}
\author{A.~Boeltzig$^{5}$}
\author{R.~Bonnel$^1$}
\author{C.~Broggini$^{10}$}
\author{C. G.~Bruno$^1$}
\author{A.~Caciolli$^{10,11}$}
\author{M.\,Campostrini$^{14}$}
\author{F.~Cavanna$^{15}$}\email{francesca.cavanna@to.infn.it}
\author{T.\,Chillery$^{12}$}
\author{G.~F.~Ciani$^{9}$}
\author{P.\,Colombetti$^{15,16}$}
\author{A.\,Compagnucci$^{12,13}$}
\author{P.~Corvisiero$^{3,4}$}
\author{L.~Csedreki$^{17}$}
\author{T.~Davinson$^1$}
\author{R.~Depalo$^{18,19}$}
\author{A.~Di Leva$^{6,7}$}
\author{A.\,D'Ottavi$^{18,19}$}
\author{Z.~Elekes$^{17,20}$}
\author{F.~Ferraro$^{12}$}
\author{A.~Formicola$^{21}$}
\author{Zs.~F\"ul\"op$^{17}$}
\author{G.~Gervino$^{15,16}$}
\author{R.~Gesu\`e$^{12,13}$}
\author{G.\,Gosta$^{18,19}$}
\author{A.~Guglielmetti$^{18,19}$}
\author{C.~Gustavino$^{21}$}
\author{Gy.~Gy\"urky$^{17}$}
\author{G.~Imbriani$^{6,7}$}
\author{J.~Jones$^1$}
\author{J.~Jos\'e$^{22, 23}$}
\author{M.~Junker$^{12}$}
\author{M.~Lugaro$^{24}$}
\author{P.~Marigo$^{10,11}$}
\author{J.~Marsh$^1$}
\author{R.~Menegazzo$^{10}$}
\author{D.~Mercogliano$^{6,7}$}
\author{D.~Piatti$^{10,11}$}
\author{P.~Prati$^{3,4}$}
\author{D.\,Rapagnani$^{6,7}$}
\author{V.\,Rigato$^{14}$}
\author{D.\,Robb$^1$}
\author{M.\,Rossi$^{3,4}$}
\author{R.\,Sariyal$^{15}$}
\author{J.~Skowronski$^{10,11}$}
\author{O.~Straniero$^{7,25}$}
\author{T.~Sz\"ucs$^{17}$}
\author{S.\,Turkat$^{10,11}$}
\author{M.\,Vagnoni$^{21}$}
\author{S.~Zavatarelli$^{3}$}\email{zavatare@ge.infn.it}

\affiliation{    $~$}
%\affiliation{$^a$INFN, Bari, Italy}

\affiliation{$^1$School of Physics and Astronomy, University of Edinburgh, EH9 3FD Edinburgh, United Kingdom}

\affiliation{$^2$School of Mathematics and Physics, University of Surrey, Guildford, GU2 7XH, United Kingdom}

\affiliation{$^3$Istituto Nazionale di Fisica Nucleare (INFN), Sezione di Genova, Via Dodecaneso 33, 16146 Genova, Italy}

\affiliation{$^{4}$Universit\`a degli Studi di Genova, Via Dodecaneso 33, 16146 Genova, Italy}

\affiliation{$^5$Helmholtz-Zentrum Dresden-Rossendorf, Dresden, Germany}

\affiliation{$^6$Universit\`a degli Studi di Napoli "Federico II", Dipartimento di Fisica "E. Pancini", Via Cintia, 80126 Napoli, Italy}

\affiliation{$^7$INFN, Sezione di Napoli, Via Cintia, 80126 Napoli, Italy}

\affiliation{$^8$Universit\`a degli Studi di Bari, 70125 Bari, Italy}
\affiliation{$^{9}$INFN, Sezione di Bari, 70125 Bari, Italy}

\affiliation{$^{10}$INFN, Sezione di Padova, Via F. Marzolo 8, 35131 Padova, Italy}
\affiliation{$^{11}$Dipartimento di Fisica e Astronomia, Universit\`a degli Studi di Padova, Via F. Marzolo 8, 35131 Padova, Italy}

\affiliation{$^{12}$INFN, Laboratori Nazionali del Gran Sasso (LNGS), Assergi (AQ), Italy}
\affiliation{$^{13}$Gran Sasso Science Institute (GSSI), L'Aquila, Italy, Assergi (AQ), Italy}

\affiliation{$^{14}$INFN Laboratori Nazionali di Legnaro, Via dell'Universit\`a 2, 35020 Legnaro (PD), Italy}

\affiliation{$^{15}$INFN, Sezione di Torino, Via P. Giuria 1, 10125 Torino, Italy}
\affiliation{$^{16}$Università degli Studi di Torino, Torino, Italy}

\affiliation{$^{17}$HUN-REN Institute for Nuclear Research (ATOMKI), Debrecen, Hungary}

\affiliation{$^{18}$Universit\`a degli Studi di Milano, Via G. Celoria 16, 20133 Milano, Italy}
\affiliation{$^{19}$INFN, Sezione di Milano, Via G. Celoria 16, 20133 Milano, Italy}

\affiliation{$^{20}$Institute of Physics, Faculty of Science and Technology, University of Debrecen, Egyetem tér 1., H-4032 Debrecen, Hungary}

\affiliation{$^{21}$INFN, Sezione di Roma, Piazzale A. Moro 2, 00185 Roma, Italy}

\affiliation{$^{22}$Universitat Politècnica de Catalunya, 08019 Barcelona, Spain}
\affiliation{$^{23}$Institut d'Estudis Espacials de Catalunya, 08034 Barcelona, Spain}

\affiliation{$^{24}$Konkoly Observatory, Research Centre for Astronomy and Earth Sciences, Hungarian Academy of Sciences, 1121 Budapest, Hungary}

\affiliation{$^{25}$Osservatorio Astronomico di Collurania, Teramo, Italy}

\collaboration{LUNA Collaboration} 

\date{\today}% It is always \today, today,
             %  but any date may be explicitly specified

\begin{abstract}
In the nova temperature range,  0.1 GK $< T <$ 0.4 GK, several low-energy resonances dominate the \nepg reaction rate, which is currently affected by large uncertainties. 
We present a high-precision study of the resonances at $E^{\rm{lab}}_{\rm{r}}$ = 127.3, 271.4, 272.3, 291.5, and 352.6 keV, measured directly at the Laboratory for Underground Nuclear Astrophysics in Italy. 
The strengths of the 127.3, 271.4, and 291.5 keV resonances are consistent with previous measurements within 1$\sigma$. However, for the 272.3 keV and 352.6 keV resonances, we report strength values of (129.9 $\pm$ 5.8) meV and (14.9 $\pm$ 0.8) meV, respectively, more than a factor of 1.5 higher than literature values. In addition, we report on new branching ratios for the 127.3, 272.3, and 352.6 keV resonances, leading to updated decay schemes. 
Finally, we present a revised thermonuclear reaction rate and investigate its impact on the NeNa nucleosynthesis.

\end{abstract}

%\keywords{Suggested keywords}%Use showkeys class option if keyword
                              %display desired
\maketitle

%\tableofcontents

\emph{Introduction}---The neon-sodium (NeNa) cycle in hydrogen burning allows the conversion of four protons into helium, using neon and sodium isotopes as catalysts \cite{izzard2007reaction}. 
Although it contributes minimally to stellar energy generation due to the high Coulomb barriers
involved, the NeNa cycle plays a key role in the synthesis of isotopes ranging from $^{20}$Ne to $^{24}$Mg. 
Furthermore, the NeNa cycle can bridge light element nucleosynthesis towards heavier elements \cite{Marion57-ApJ}.
In recent years, significant progress has been made in reducing the uncertainties of many nuclear reactions in the NeNa cycle and in understanding their impact on the resulting abundances, thanks to the efforts of several research groups \cite{
cavannaprl2015, Depalo16-PRC, kellyprc2017, ferraro2018direct, williams2020first, masha2023prc, Takacs24-PRC}. However, the $\mathrm{^{21}Ne(p, \gamma)^{22}Na}$, and $\mathrm{^{23}Na(p, \alpha)^{20}Ne}$ reaction rates are still affected by large uncertainties. 
Therefore, high-precision experimental data are needed to improve our understanding of the production of Ne-Na isotopes in both asymptotic giant branch (AGB) stars and explosive stellar environments such as novae and supernovae.
In AGB stars, the NeNa cycle is activated at the base of the convective envelope during hot bottom burning (HBB) at $T$ \,$\leq$\, 0.1\,GK, when the stellar mass is $\geqslant$ 4\,$M_{\odot}$. 
In explosive environments, the radioactive $^{22}$Na (with a half life of 2.6 yr), plays a particularly significant role. 
It offers a direct probe of nucleosynthesis in novae events~\cite{weiss199022}, i.e., thermonuclear explosions on the surface of a white dwarf in a binary system with a less evolved companion. Its production is strongly associated with oxygen-neon (ONe) white dwarfs, the descendants of stars with initial masses in the range of 8–10~$M_\odot$.
The $^{22}$Na decay emits a 1.275 MeV photon~\cite{BASUNIA201569}, whose
detection would allow to estimate the amount of $^{22}$Na produced in an ONe nova. 
Detecting the 1.275 MeV $\gamma$-rays requires a nearby nova explosion, within a distance of a few kpc (1 kpc\,=\,3262 light-years) \cite{Fougeres:2022upo}. 
Although no $^{22}$Na $\gamma$-rays have been observed, this may be achieved by future satellite missions with advanced detector technology and unprecedented line sensitivity~\cite{fryer2019catching}. To fully harness their potential, especially in the event of a nearby nova, accurate nuclear physics inputs must be established.

Another astrophysical scenario where $^{22}$Na plays a crucial role is supernovae. In core-collapse supernovae, even though the 1.275 MeV line from $^{22}$Na~\cite{woosley2005physics,janka2007theory} is much weaker than the 1.238 MeV~\cite{JUNDE20111513} $\gamma$-ray line from the decay of $^{56}$Co, $^{22}$Na production can be traced through the signature of its decay into $^{22}$Ne, observed as a strong excess in meteoritic stardust grains~\cite{amari2008sodium}. 
This excess likely originates from $^{22}$Na trapped in grains formed during supernovae explosions~\cite{amari1990interstellar}. A precise determination of the \nepg reaction rate can help to identify the paternity of presolar meteoritic grains, through better estimates of a number of Ne isotopic ratios \cite{Amari_2001}.

In the temperature range relevant to novae, from 0.1 to 0.4 GK, the \nepg reaction rate is dominated by two low-energy resonances at $E^{\rm{lab}}_{\rm{r}}$ = 127.3 and 272.3 keV. 
At higher temperatures, $T >$ 0.4 GK, relevant to explosive nucleosynthesis during supernovae, also the $E^{\rm{lab}}_{\rm{r}}$ = 291.5 and 352.6 keV resonances contribute significantly to the reaction rate~\cite{gorres1982search}.
The strengths of all these low-energy resonances have been previously measured by~\citet{gorres1982search,berg1977proton,gorres1983proton}, and \citet{becker1992low} with a typical reported uncertainty of 15-20$\%$~\cite{gorres1982search,becker1992low}.  
In this work, we report on new direct and precise measurements of the resonances at $E^{\rm{lab}}_{\rm{r}}$ = 127.3, 271.4, 272.3, 291.5, and 352.6 keV.

\begin{figure}
\centering
\includegraphics[width=\linewidth]{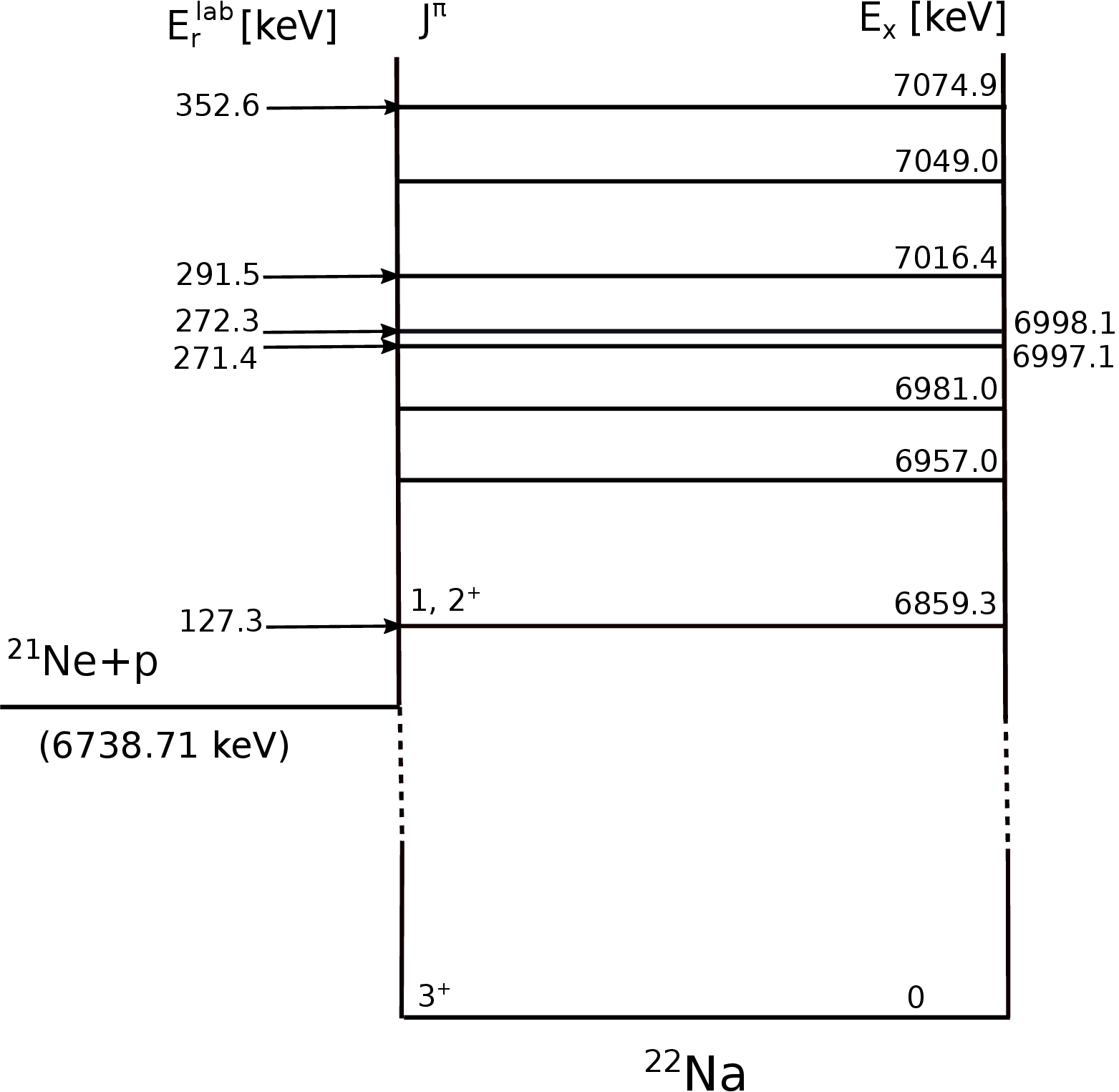}
\caption{
Partial level scheme of $^{22}$Na up to the energy
$E_x$ = 7074.9 keV. The
excited states and corresponding resonance energies $E^{\rm{lab}}_{\rm{r}}$ investigated
in the present experiment are indicated.
}
\label{figure-1}     
\end{figure}

\emph{Experiment}---The \nepg reaction (\textit{Q}-value = 6738.71 keV~\cite{BASUNIA201569}) was studied at the Laboratory for Underground Nuclear Astrophysics (LUNA) \cite{aliotta2022exploring}, taking advantage of the low environmental background levels \cite{Caciolli09-EPJA,Szucs10-EPJA} at the Gran Sasso National Laboratory (LNGS), Italy.
The experiment was performed using the high-intensity proton beam (beam current up to 300 $\mu$A) provided by the LUNA 400 kV accelerator. 
The beam entered a parallelepiped-shaped windowless reaction chamber filled with 2 mbar pressure of either natural neon ($^{20}$Ne = 90.48$\%$, $^{21}$Ne = 0.27$\%$, and $^{22}$Ne = 9.25$\%$) or $^{21}$Ne enriched neon ($^{20}$Ne = 5.53$\%$, $^{21}$Ne = 59.1$\%$, and $^{22}$Ne = 35.37$\%$) gas, both used in recirculation mode \cite{cavanna2014epja}. 
The beam was stopped on a Cu calorimeter that measured its beam current \cite{ferraro2018epja}.  
The $\gamma$-rays emitted by the \nepg reaction were detected using two large high-purity germanium (HPGe) detectors with relative detection efficiencies of 90$\%$ and 130$\%$, mounted at two different positions along the beam path. The entire setup was surrounded by a 20 cm lead shielding to further reduce the environmental background.

The $\gamma$-detection efficiency of the detectors was characterized using calibrated radioactive sources ($^{133}$Ba, $^{137}$Cs, and $^{60}$Co) for the low energy region (up to 1.3\,MeV), and the well-known $^{14}$N(p,$\gamma$)$^{15}$O resonance at $E_{\rm r}^{\rm lab}$ = 278 keV \cite{daigle2016} to extend the efficiency determination up to 7\,MeV. The adopted experimental setup was implemented in the LUNA GEANT4 framework--\texttt{SimLUNA} and was fine-tuned using experimental data. Description of setup and Monte Carlo simulations can be found in~\cite{masha2024epja}.

\begin{figure}
\centering
\includegraphics[width=1.1\linewidth]{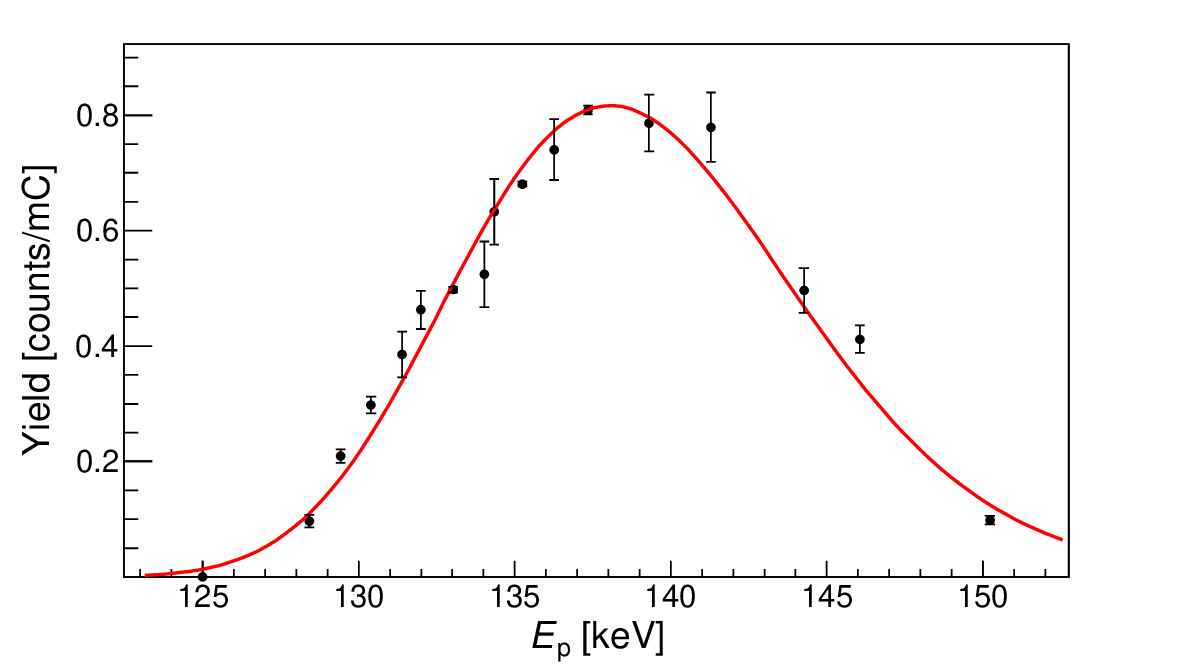}
%{scan_126.eps}
\caption{Experimental scan of the $E^{\rm{lab}}_{\rm{r}}$ = 127.3 keV resonance as a function of the proton beam energy, measured with the HPGe 130\% detector. A fit to the experimental data is also shown.}
\label{fig_scan126keV}     
\end{figure}

\emph{Data taking and analysis}---Figure~\ref{figure-1} shows the level scheme of $^{22}$Na in the energy range accessible with the LUNA 400 kV accelerator. For each measured resonance, as a first step, a run was performed at the beam energy where the maximum yield was expected according to the literature resonance energy, taking into account the beam energy loss and the efficiency profile of the detectors. 
A resonance scan was then performed by varying the beam energy in steps of 1--2 keV and measuring the yield of one of the most intense primary $\gamma$ transitions (i.e., transitions from the resonance under study
to a given state in $^{22}$Na). 
Figure~\ref{fig_scan126keV} shows the scan of the resonance at $E^{\rm{lab}}_{\rm{r}}$ = 127.3 keV, measured with the HPGe 130\% detector. 
%is shown in Fig.~\ref{fig_scan126keV}.
Scans for the remaining resonances are provided in the Supplemental Material~\cite{Sidhu_supplement}. 

Using these excitation functions, the resonance energies for the five resonances were determined. The details of the adopted method and the results are reported in~\cite{masha2024epja}.

\begin{figure*}
\centering
\includegraphics[width=0.85\linewidth]{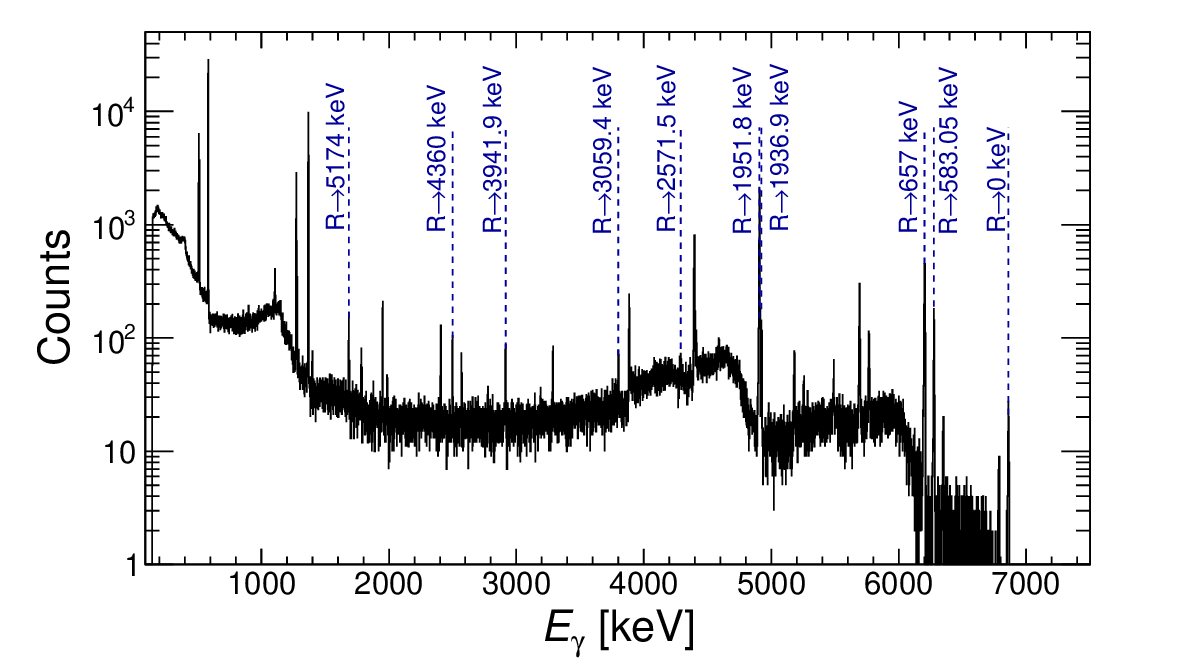}
\caption{Experimental spectrum measured with the HPGe 130\% detector at the maximum detection yield for the $E^{\rm{lab}}_{\rm{r}}$ = 127.3 keV resonance. 
Primary transitions from the $E_x$ = 6859.3 keV excited state (R) are labeled.}
\label{figure-2}     
\end{figure*}

\begin{figure*}
\centering
\includegraphics[width=0.85\linewidth]{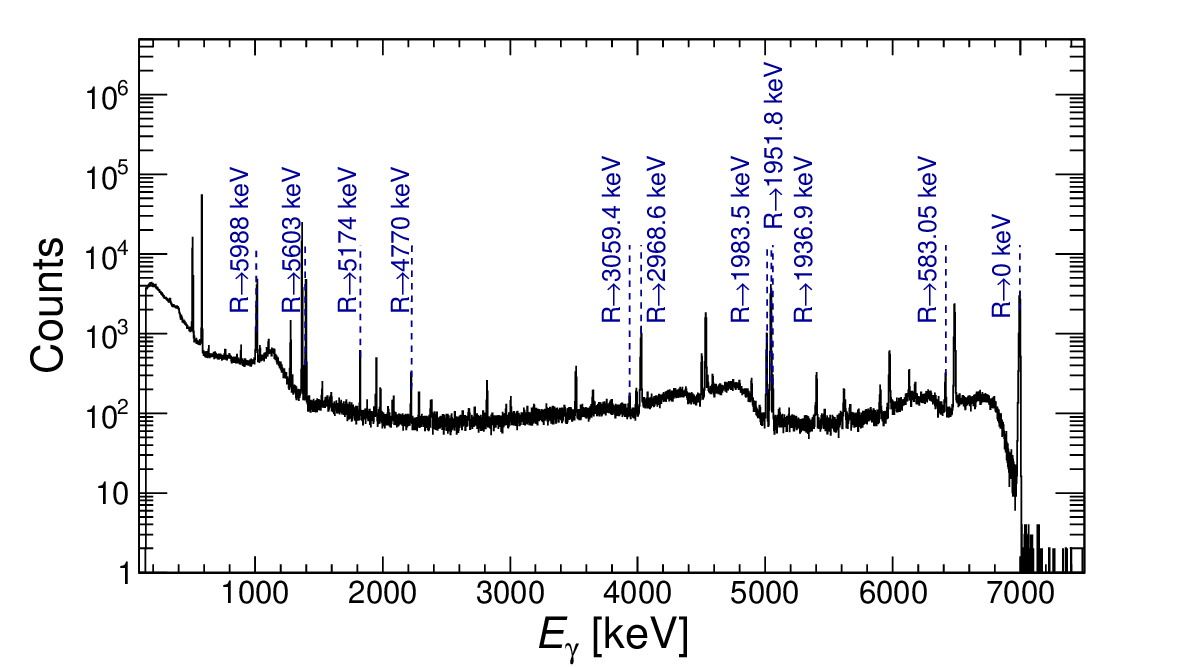}
\caption{Experimental spectrum measured with the HPGe 130\% detector at the maximum detection yield of the $E^{\rm{lab}}_{\rm{r}} = 272.3$ keV resonance.
Primary transitions from the excited state (R) at $E_x = 6998.1$ keV are labeled.}
\label{figure-3}     
\end{figure*}

To measure the resonance strengths, long runs (up to 50 C) were performed at beam energies where the maximum yield was observed, corresponding to the resonance being populated in front of the two HPGe detectors.
For the $E^{\rm{lab}}_{\rm{r}}$ = 127.3, 272.3, and 291.5 keV resonances, new primary $\gamma$ transitions were discovered in addition to those already reported in literature~\cite{BASUNIA201569}. 
Figures~\ref{figure-2} and~\ref{figure-3} display the experimental spectra 
obtained for the resonances at 127.3 and 272.3 keV, respectively. Primary $\gamma$ transitions are marked; the corresponding branching ratios (BRs) are provided in Table~\ref{tab:table0}.
The experimental spectrum and the BRs of the $E^{\rm{lab}}_{\rm{r}}$ = 291.5 keV resonance are reported in~\cite{masha2024epja}.
For the 352.6 keV resonance, the intensity of the observed primary $\gamma$ transitions (Table~\ref{tab:table0}) is in agreement with the values reported in literature~\cite{BASUNIA201569}.

\begin{table*}\caption{Branching ratios (BRs) for the primary gammas from the levels at $E_{x}$\,=\,7074.9, 6998.1 and 6859.3\,keV corresponding to the resonances at $E^{\rm{lab}}_{\rm{r}}$ = 352.6, 272.3 and 127.3 keV, respectively. 
The format of BRs evaluated in this work is value $\pm$ total uncertainty (=\, $\sqrt{\rm{statistical\,  uncertainty^2+systematic\,uncertainty^2}}$), both in \%.
For comparison, the branching ratios from the literature~\cite{BASUNIA201569}, along with their total uncertainties, are also provided.} 
\centering
\begin{tabular}{ccccccc}
\hline\hline   
\multirow{2}[3]{*}{$E_{x}$ (keV)} & \multicolumn{2}{c}{7074.9} & \multicolumn{2}{c}{6998.1}  & \multicolumn{2}{c}{6859.3} \\
\cmidrule(lr){2-3} \cmidrule(lr){4-5} \cmidrule(lr){6-7}
 & LUNA & lit. value & LUNA & lit. value & LUNA & lit. value  \\  \hline 
5995	&	5.4  $\pm$ 0.3 	&	3.64 $\pm$ 0.38  	&		&		&		&		\\
5988	&		&		&	2.4 $\pm$ 0.2 	&	---	&		&		\\
5959	&	12.6  $\pm$ 0.7 	&	9.71 $\pm$ 0.64  	&		&		&		&		\\
5739	&	0.58 $\pm$ 0.04 	&	0.61 $\pm$ 0.17  	&		&		&		&		\\
5725	&	0.41 $\pm$ 0.04 	&	0.36  $\pm$ 0.08 	&		&		&		&		\\
5700	&	2.7 $\pm$ 0.2 	&	14.71  $\pm$ 1.74 	&		&		&		&		\\
5603	&		&		&	0.5 $\pm$ 0.2  	&	---	&		&		\\
5174	&	4.6 $\pm$ 0.3 	&	3.64  $\pm$ 0.28 	&	0.98 $\pm$  0.06 	&	0.96 $\pm$ 0.22 	&	1.25 $\pm$ 0.09 	&	---	\\
5062.5	&	0.75 $\pm$ 0.05 	&	0.53  $\pm$ 0.17 	&		&		&		&		\\
4770	&		&		&	0.66 $\pm$  0.04 	&	---	&		&		\\
4622	&	0.57 $\pm$ 0.05 	&	0.53  $\pm$ 0.17 	&		&		&		&		\\
4583	&	0.67 $\pm$ 0.05 	&	0.61  $\pm$ 0.17 	&		&		&		&		\\
4360	&	0.56 $\pm$ 0.05 	&	0.61  $\pm$ 0.17 	&		&		&	1.16 $\pm$ 0.09	&	---	\\
4296.2	&	0.50 $\pm$ 0.05 	&	0.69  $\pm$ 0.17 	&		&		&		&		\\
3941.9	&	0.90 $\pm$ 0.06 	&	0.69  $\pm$ 0.25	&		&		&	0.98 $\pm$ 0.08	&	---	\\
3519.1	&	1.31 $\pm$ 0.09 	&	1.30  $\pm$ 0.17	&		&		&		&		\\
3059.4	&	3.1 $\pm$ 0.2 	&	2.16  $\pm$ 0.26	&	0.38 $\pm$ 0.05	&	---	&	1.1 $\pm$ 0.1	&	---	\\
2968.6	&		&		&	6.0 $\pm$  0.4	&	4.48  $\pm$ 0.97	&		&		\\
2571.5	&		&		&		&		&	0.6 $\pm$ 0.1	&	---	\\
1983.5	&		&		&	6.8 $\pm$  0.4	&	6.30  $\pm$ 0.57	&		&		\\
1951.8	&	5.7  $\pm$ 0.4 	&	5.83  $\pm$ 1.40	&	33.2  $\pm$ 1.9	&	30.95  $\pm$ 2.37	&	64.2 $\pm$  3.8	&	80  $\pm$ 13	\\
1936.9	&		&		&	1.5  $\pm$ 0.1	&	2.03  $\pm$ 0.54	&	3.2 $\pm$ 0.2	&	---	\\
657	&	27.7  $\pm$ 1.6 	&	27.75  $\pm$ 1.90	&		&		&	19.6  $\pm$ 1.2	&	20  $\pm$ 9	\\
583.05	&	31.9  $\pm$ 1.8 	&	26.09  $\pm$ 1.88	&	2.2 $\pm$ 0.2	&	1.92  $\pm$ 0.43	&	6.8 $\pm$  0.4	&	---	\\
0	&	---	&	0.53  $\pm$ 0.17	&	45.3 $\pm$ 2.6	&	53.36  $\pm$ 2.76	&	1.16 $\pm$ 0.09	&	---	\\

\hline\hline   
\end{tabular}
\label{tab:table0}
\end{table*}

The intrinsic widths, $\Gamma$, of all the resonances investigated in this work are significantly smaller (less than a few eV) than the beam energy loss inside the target ($\Gamma \ll \Delta E_{\rm c.m.}$) \cite{becker1992low} and, hence, the thick-target condition~\cite{RevModPhys.20.236} is valid to calculate the resonance strength $\omega \gamma$.
Under this assumption, the $\omega \gamma$ is determined as \cite{rolfs1988cauldrons}:
\begin{equation}
    \omega \gamma = \frac{2}{\lambda_{\rm r}^2} Y \epsilon_{\rm r} \frac{M}{m+M},
\label{equation-1}
\end{equation}
where $\lambda_{\rm r}$ is the de Broglie wavelength at the resonance energy, $\epsilon_{\rm r} $ is the effective stopping power in the laboratory system, and $m$ and $M$ are
projectile and target masses, respectively. The total yield, $Y$, is obtained as:
\begin{equation}
    Y=\frac{1}{Q}\sum_i\frac{N_i}{\eta_i},
\end{equation}
where $Q$ is the integrated beam charge, $N_i$ is the number of counts measured for each primary transition \textit{i}, and $\eta_i$ is the corresponding full energy detection efficiency, calculated from \texttt{SimLUNA}. 
By summing over all primary $\gamma$ transitions, the total resonance strength is determined. The effect of true coincidence summing on $\gamma$-ray cascades and the influence of beam energy straggling are included in the simulation~\cite{masha2024epja,bemmerer2018effect}. 

For the  $E^{\rm{lab}}_{\rm{r}}$ = 271.4 keV resonance ($E_{x}$ = 6997.1 keV, see Fig.~4 in~\cite{Sidhu_supplement}), it was not possible to calculate the resonance strength from the total yield \textit{Y}, as the 5013.4 keV and 6995.9 keV primary $\gamma$ rays are mixed with those from the $E^{\rm{lab}}_{\rm{r}}$ = 272.3 keV resonance ($E_{x}$ = 6998.1 keV) just 1 keV apart. 
In addition, according to the literature, the 5468.7 keV primary $\gamma$ is weak (BR = 4.9 $\pm$ 1.0$\%$~\cite{BASUNIA201569}), and its signal was not observed in the present experimental data. 
Therefore, the resonance strength was determined using the 2287 keV primary $\gamma$, normalized to its BR.

For the  $E^{\rm{lab}}_{\rm{r}}$ = 352.6 keV resonance, all primary $\gamma$ transitions, except the 7075.4 keV primary $\gamma$ (which is mixed with a broad energy bump, see Fig.~5 in~\cite{Sidhu_supplement}), were used to determine the~\textit{Y}.
This primary $\gamma$ should have very low intensity (BR = (0.53 $\pm$ 0.17)$\%$~\cite{BASUNIA201569}) according to the literature, and its contribution is thus included in the error budget.  

%The resonance strengths for the measured resonances are given in Table~\ref{table-3}. The strengths of the $E^{\rm{lab}}_{\rm{r}}$ = 127.3 keV, 272.3 keV, and 291.5 keV resonances have been determined with $\sim$4\% uncertainty, whereas, the strengths of the $E^{\rm{lab}}_{\rm{r}}$ = 271.4 keV and 352.6 keV resonances have been determined with $\sim$19\% and $\sim$5\% uncertainty, respectively.

For the $E^{\rm{lab}}_{\rm{r}}$ = 127.3 keV, 271.4 keV, and 291.5 keV resonances, the $\omega \gamma$ agrees with the measurements reported in~\citet{gorres1982search} and \citet{becker1992low} within 1$\sigma$ uncertainty.
Instead, for the $E^{\rm{lab}}_{\rm{r}}$ = 272.3 keV and 352.6 keV resonances, the measured $\omega \gamma$ shows a 3.4$\sigma$ and 4.3$\sigma$ deviation, respectively, from the $\omega \gamma$ values reported  in~\citet{gorres1982search}.

For each resonance, we calculated the electron screening correction factor \textit{f} in the adiabatic approximation~\cite{assenbaum1987effects}; the results are reported in the last column of Table \ref{table-3}. 
We underline that, except for the $E^{\rm{lab}}_{\rm{r}}$ = 127.3 keV resonance, the correction is smaller than the quoted uncertainties on the strength. 
In addition, according to a recent paper from~\citet{iliadis2023laboratory}, the electron screening correction for narrow resonances should always be neglected.

\begin{table}%[H] add [H] placement to break table across pages
\caption{\label{tab:1} Resonance strengths measured in this work (not corrected for electron screening) and literature values \cite{gorres1982search,becker1992low}. The format of LUNA data is value $\pm$ statistical uncertainty $\pm$ systematic uncertainty. In the last column, the electron screening factor \textit{f} is reported. }
\begin{ruledtabular}
\begin{tabular}{c c c c }
%$E_{\rm{r}}^{\rm lab}$ [keV] &  $\omega \gamma_{\rm literature} $ [meV] & $\omega \gamma_{\rm LUNA}$ [meV] & \textit{f} \\
$E_{\rm{r}}^{\rm lab}$  &  $\omega \gamma_{\rm literature}\footnote{Values taken from Table 1 of \cite{gorres1982search}, except for the resonance strength of $E_{\rm{r}}^{\rm lab}$\,=127.3\,keV taken from Table 1 of \cite{becker1992low}.} $  & $\omega \gamma_{\rm LUNA}$  & \textit{f} \\
{[keV]} & {[meV]} & {[meV]} & \\ \hline
127.3 & 0.0375 $\pm$ 0.0075 \cite{becker1992low} &  0.0375 $\pm$ 0.0002 $\pm$ 0.0017 &  1.10  \\
271.4 &  2.125 $\pm$ 0.375 \cite{gorres1982search}  &  2.7 $\pm$ 0.3 $\pm$ 0.4 &  1.03 \\
272.3 & 82.5 $\pm$ 12.5 \cite{gorres1982search} & 129.9 $\pm$ 0.4 $\pm$ 5.8 & 1.03 \\
291.5 &  2.00 $\pm$ 0.37 \cite{gorres1982search} &  1.99 $\pm$ 0.01 $\pm$ 0.09 &  1.03\\ 
352.6 &  8.125 $\pm$ 1.375 \cite{gorres1982search} &  14.9 $\pm$ 0.4 $\pm$ 0.7 &  1.02  \\ [1ex]
% Lines of table here ending with \\
\end{tabular}
\label{table-3}
 \end{ruledtabular}
\end{table}

\emph{Thermonuclear reaction rates and implications}---We have updated the thermonuclear reaction rates for the \nepg reaction using the resonance strengths reported in this work. The rate is calculated using the Monte Carlo code \texttt{RATESMC} \cite{sallaska2013starlib}. For the five low-energy resonances at $E^{\rm{lab}}_{\rm{r}}$ = 127.3, 271.4, 272.3, 291.5, 352.6 keV, we adopted our new measured values. For all other resonances, we used the data from \citet{iliadis2010charged}.
The total reaction rate in the temperature range from 0.01 to 10 GK is provided in Table I in the Supplemental Material~\cite{Sidhu_supplement}.

\begin{figure*}
\centering
\includegraphics[width=0.7\linewidth]{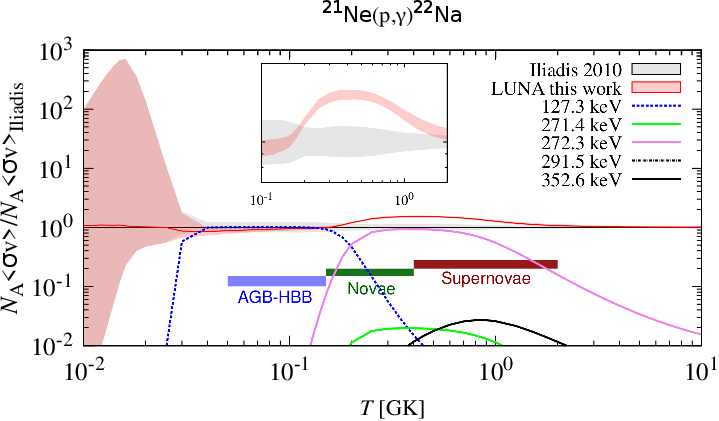}
\caption{LUNA thermonuclear reaction rate for the \nepg reaction (red) compared with Iliadis {\it et al.} \cite{ILIADIS2010_II} (grey). The shaded area shows the $1\mathrm{\sigma}$ uncertainty. The rates are normalized to Iliadis {\it et al.} \cite{ILIADIS2010_II}. %The top x-axis shows the Gamow energy for the given temperature range. 
The AGB, nova, and supernova temperature regions of interest are also shown.}
\label{rate}     
\end{figure*}

A comparison between the updated LUNA rate and the rate from Iliadis {\it et al.} is shown in Fig.~\ref{rate}. The new rate is in agreement with the previous one \cite{ILIADIS2010_II}, except in the temperature range relevant for classical novae (0.1–0.4 GK), where the rate is dominated by the $E^{\rm{lab}}_{\rm{r}}$ = 272.3 keV resonance. In this range, the LUNA rate is approximately 23\% higher than the one reported in Iliadis {\it et al.} \cite{ILIADIS2010_II}. At lower temperatures ($T <$ 0.1 GK), relevant for AGB stars, the new rate is compatible with the previous one, but more precise.

To evaluate the astrophysical impact of the updated rate, we used different stellar evolution codes according to the astrophysical scenario. We performed hydrodynamic simulations of ONe novae using the \texttt{SHIVA} code~\cite{Jose_1998,Jose16-Book}, which is widely applied in modeling stellar explosions such as classical novae, X-ray bursts, and Type Ia supernovae. Simulations were performed for three different white dwarf masses: 1.05, 1.15, and 1.25 $M_\odot$. For each case, two models were computed, using the new LUNA recommended \nepg reaction rate and the Iliadis {\it et al.} reaction rate~\cite{ILIADIS2010_II}. Further calculations were performed using the LUNA \nepg high and low rates (1$\sigma$ uncertainty). 
For both tests, the results show that the mass-averaged abundances of isotopes sensitive to the \nepg reaction, including $^{22}$Ne, $^{22}$Na, and $^{25}$Mg, remain essentially unchanged, due to the well constrained reaction rate in the LUNA energy range.

For the AGB scenario, we considered a $5\,M_\odot$ star with solar metallicity ($Z = 0.014$), representative of HBB conditions. Nucleosynthesis calculations using both the LUNA and Iliadis {\it et al.}  \cite{ILIADIS2010_II} rates showed no significant differences in the final elemental abundances. This confirms the limited sensitivity of the $^{21}$Ne(p,$\gamma$)$^{22}$Na reaction rate in AGB stars and further supports the robustness of the thermonuclear reaction rate.

\emph{Conclusions}---We have presented new high-precision direct measurements of five low-energy resonances in the \nepg reaction performed at the LUNA underground laboratory.
The resonance strengths at $E^{\rm{lab}}_{\rm{r}}$ = 127.3, 272.3, and 291.5 keV were determined with uncertainties as low as 4\%, while those at 271.4 and 352.6 keV were measured with uncertainties of 19\% and 5\%, respectively.
In particular, the strength of the 272.3 keV resonance, which affects the nova energy range, shows a 3.4$\sigma$ increase compared to previous measurements. 
New branching ratios were determined for three resonances ($E^{\rm{lab}}_{\rm{r}}$ = 127.3,  272.3, and 352.6 keV), providing updated decay schemes.
The improved resonance strengths were used to determine a new thermonuclear reaction rate using Monte Carlo techniques and to investigate the impact on the NeNa nucleosynthesis. The new rate is about 23\% higher than previous evaluations for temperatures near peak values for novae, mainly due to the new data for the 272.3 keV resonance. Nevertheless, the hydrodynamic models of the classical novae show no variation on $^{22}$Na production in novae, when this new rate is used. In AGB stars, the impact of the revised rate is negligible, confirming the limited sensitivity of that scenario.

\emph{Acknowledgments}---D.C. and the technical staff of the LNGS are gratefully acknowledged for their support.
We acknowledge funding from: INFN,
the European Union (ERC-CoG STARKEY, no. 615604; ERC-StG SHADES, no. 852016; and ChETEC-INFRA, no. 101008324),
Deut\-sche For\-schungs\-ge\-mein\-schaft (DFG, BE~4100-4/1),
the Helm\-holtz Association (ERC-RA-0016),
the Hungarian National Research, Development, and Innovation Office (NKFIH K134197),
the European Collaboration for Science and Technology (COST Action ChETEC, CA16117). E.M. acknowledges an Alexander von Humboldt postdoctoral fellowship.
M.A., C.G.B., T.D., and R.S.S. acknowledge funding from STFC (grant ST/P004008/1).
J.J. acknowledges support from the Spanish MINECO grant PID2023-148661NB-I00, the E.U. FEDER funds,  and the AGAUR/Generalitat de Catalunya grant SGR-386/2021.
We acknowledge financial support under the National Recovery and Resilience Plan (NRRP), Mission 4, Component 2, Investment 1.1, Call for tender No. 104 published on 2.2.2022 by the Italian Ministry of University and Research (MUR), funded by the European Union – NextGenerationEU– Projects Title CaBS – CUP E53D230023 - and Title SOCIAL – CUP I53D23000840006 - Grant Assignment Decree No. 974 adopted on 30/06/2023 by the Italian Ministry of Ministry of University and Research (MUR).
For the purpose of open access, authors have applied a Creative Commons Attribution (CC BY) license to any Author Accepted Manuscript version arising from this submission.
%The authors declare that they have no known competing financial interests or personal relationships that could have appeared to influence the work reported in this Letter.

\bibliography{references.bib}

%apsrev4-2.bst 2019-01-14 (MD) hand-edited version of apsrev4-1.bst
%Control: key (0)
%Control: author (8) initials jnrlst
%Control: editor formatted (1) identically to author
%Control: production of article title (0) allowed
%Control: page (0) single
%Control: year (1) truncated
%Control: production of eprint (0) enabled
\begin{thebibliography}{41}%
\makeatletter
\providecommand \@ifxundefined [1]{%
 \@ifx{#1\undefined}
}%
\providecommand \@ifnum [1]{%
 \ifnum #1\expandafter \@firstoftwo
 \else \expandafter \@secondoftwo
 \fi
}%
\providecommand \@ifx [1]{%
 \ifx #1\expandafter \@firstoftwo
 \else \expandafter \@secondoftwo
 \fi
}%
\providecommand \natexlab [1]{#1}%
\providecommand \enquote  [1]{``#1''}%
\providecommand \bibnamefont  [1]{#1}%
\providecommand \bibfnamefont [1]{#1}%
\providecommand \citenamefont [1]{#1}%
\providecommand \href@noop [0]{\@secondoftwo}%
\providecommand \href [0]{\begingroup \@sanitize@url \@href}%
\providecommand \@href[1]{\@@startlink{#1}\@@href}%
\providecommand \@@href[1]{\endgroup#1\@@endlink}%
\providecommand \@sanitize@url [0]{\catcode `\\12\catcode `\$12\catcode `\&12\catcode `\#12\catcode `\^12\catcode `\_12\catcode `\%12\relax}%
\providecommand \@@startlink[1]{}%
\providecommand \@@endlink[0]{}%
\providecommand \url  [0]{\begingroup\@sanitize@url \@url }%
\providecommand \@url [1]{\endgroup\@href {#1}{\urlprefix }}%
\providecommand \urlprefix  [0]{URL }%
\providecommand \Eprint [0]{\href }%
\providecommand \doibase [0]{https://doi.org/}%
\providecommand \selectlanguage [0]{\@gobble}%
\providecommand \bibinfo  [0]{\@secondoftwo}%
\providecommand \bibfield  [0]{\@secondoftwo}%
\providecommand \translation [1]{[#1]}%
\providecommand \BibitemOpen [0]{}%
\providecommand \bibitemStop [0]{}%
\providecommand \bibitemNoStop [0]{.\EOS\space}%
\providecommand \EOS [0]{\spacefactor3000\relax}%
\providecommand \BibitemShut  [1]{\csname bibitem#1\endcsname}%
\let\auto@bib@innerbib\@empty
%</preamble>
\bibitem [{\citenamefont {Izzard}\ \emph {et~al.}(2007)\citenamefont {Izzard}, \citenamefont {Lugaro}, \citenamefont {Karakas}, \citenamefont {Iliadis},\ and\ \citenamefont {van Raai}}]{izzard2007reaction}%
  \BibitemOpen
  \bibfield  {author} {\bibinfo {author} {\bibfnamefont {R.~G.}\ \bibnamefont {Izzard}}, \bibinfo {author} {\bibfnamefont {M.}~\bibnamefont {Lugaro}}, \bibinfo {author} {\bibfnamefont {A.~I.}\ \bibnamefont {Karakas}}, \bibinfo {author} {\bibfnamefont {C.}~\bibnamefont {Iliadis}},\ and\ \bibinfo {author} {\bibfnamefont {M.}~\bibnamefont {van Raai}},\ }\href {https://doi.org/10.1051/0004-6361:20066903} {\bibfield  {journal} {\bibinfo  {journal} {Astronomy \& Astrophysics}\ }\textbf {\bibinfo {volume} {466}},\ \bibinfo {pages} {641} (\bibinfo {year} {2007})}\BibitemShut {NoStop}%
\bibitem [{\citenamefont {{Marion}}\ and\ \citenamefont {{Fowler}}(1957)}]{Marion57-ApJ}%
  \BibitemOpen
  \bibfield  {author} {\bibinfo {author} {\bibfnamefont {J.~B.}\ \bibnamefont {{Marion}}}\ and\ \bibinfo {author} {\bibfnamefont {W.~A.}\ \bibnamefont {{Fowler}}},\ }\href {https://doi.org/10.1086/146296} {\bibfield  {journal} {\bibinfo  {journal} {Astrophysical Journal}\ }\textbf {\bibinfo {volume} {125}},\ \bibinfo {pages} {221} (\bibinfo {year} {1957})}\BibitemShut {NoStop}%
\bibitem [{\citenamefont {Cavanna}\ \emph {et~al.}(2015)\citenamefont {Cavanna}, \citenamefont {Depalo}, \citenamefont {Aliotta}, \citenamefont {Anders}, \citenamefont {Bemmerer}, \citenamefont {Best}, \citenamefont {Boeltzig}, \citenamefont {Broggini}, \citenamefont {Bruno}, \citenamefont {Caciolli} \emph {et~al.}}]{cavannaprl2015}%
  \BibitemOpen
  \bibfield  {author} {\bibinfo {author} {\bibfnamefont {F.}~\bibnamefont {Cavanna}}, \bibinfo {author} {\bibfnamefont {R.}~\bibnamefont {Depalo}}, \bibinfo {author} {\bibfnamefont {M.}~\bibnamefont {Aliotta}}, \bibinfo {author} {\bibfnamefont {M.}~\bibnamefont {Anders}}, \bibinfo {author} {\bibfnamefont {D.}~\bibnamefont {Bemmerer}}, \bibinfo {author} {\bibfnamefont {A.}~\bibnamefont {Best}}, \bibinfo {author} {\bibfnamefont {A.}~\bibnamefont {Boeltzig}}, \bibinfo {author} {\bibfnamefont {C.}~\bibnamefont {Broggini}}, \bibinfo {author} {\bibfnamefont {C.~G.}\ \bibnamefont {Bruno}}, \bibinfo {author} {\bibfnamefont {A.}~\bibnamefont {Caciolli}}, \emph {et~al.} (\bibinfo {collaboration} {The LUNA Collaboration}),\ }\href {https://doi.org/10.1103/PhysRevLett.115.252501} {\bibfield  {journal} {\bibinfo  {journal} {Physical Review Letters}\ }\textbf {\bibinfo {volume} {115}},\ \bibinfo {pages} {252501} (\bibinfo {year} {2015})}\BibitemShut {NoStop}%
\bibitem [{\citenamefont {Depalo}\ \emph {et~al.}(2016)\citenamefont {Depalo}, \citenamefont {Cavanna}, \citenamefont {Aliotta}, \citenamefont {Anders}, \citenamefont {Bemmerer}, \citenamefont {Best}, \citenamefont {Boeltzig}, \citenamefont {Broggini}, \citenamefont {Bruno}, \citenamefont {Caciolli} \emph {et~al.}}]{Depalo16-PRC}%
  \BibitemOpen
  \bibfield  {author} {\bibinfo {author} {\bibfnamefont {R.}~\bibnamefont {Depalo}}, \bibinfo {author} {\bibfnamefont {F.}~\bibnamefont {Cavanna}}, \bibinfo {author} {\bibfnamefont {M.}~\bibnamefont {Aliotta}}, \bibinfo {author} {\bibfnamefont {M.}~\bibnamefont {Anders}}, \bibinfo {author} {\bibfnamefont {D.}~\bibnamefont {Bemmerer}}, \bibinfo {author} {\bibfnamefont {A.}~\bibnamefont {Best}}, \bibinfo {author} {\bibfnamefont {A.}~\bibnamefont {Boeltzig}}, \bibinfo {author} {\bibfnamefont {C.}~\bibnamefont {Broggini}}, \bibinfo {author} {\bibfnamefont {C.~G.}\ \bibnamefont {Bruno}}, \bibinfo {author} {\bibfnamefont {A.}~\bibnamefont {Caciolli}}, \emph {et~al.} (\bibinfo {collaboration} {LUNA Collaboration}),\ }\href {https://doi.org/10.1103/PhysRevC.94.055804} {\bibfield  {journal} {\bibinfo  {journal} {Physical Review C}\ }\textbf {\bibinfo {volume} {94}},\ \bibinfo {pages} {055804} (\bibinfo {year} {2016})}\BibitemShut {NoStop}%
\bibitem [{\citenamefont {Kelly}\ \emph {et~al.}(2017)\citenamefont {Kelly}, \citenamefont {Champagne}, \citenamefont {Downen}, \citenamefont {Dermigny}, \citenamefont {Hunt}, \citenamefont {Iliadis},\ and\ \citenamefont {Cooper}}]{kellyprc2017}%
  \BibitemOpen
  \bibfield  {author} {\bibinfo {author} {\bibfnamefont {K.~J.}\ \bibnamefont {Kelly}}, \bibinfo {author} {\bibfnamefont {A.~E.}\ \bibnamefont {Champagne}}, \bibinfo {author} {\bibfnamefont {L.~N.}\ \bibnamefont {Downen}}, \bibinfo {author} {\bibfnamefont {J.~R.}\ \bibnamefont {Dermigny}}, \bibinfo {author} {\bibfnamefont {S.}~\bibnamefont {Hunt}}, \bibinfo {author} {\bibfnamefont {C.}~\bibnamefont {Iliadis}},\ and\ \bibinfo {author} {\bibfnamefont {A.~L.}\ \bibnamefont {Cooper}},\ }\href {https://doi.org/10.1103/PhysRevC.95.015806} {\bibfield  {journal} {\bibinfo  {journal} {Physical Review C}\ }\textbf {\bibinfo {volume} {95}},\ \bibinfo {pages} {015806} (\bibinfo {year} {2017})}\BibitemShut {NoStop}%
\bibitem [{\citenamefont {Ferraro}\ \emph {et~al.}(2018{\natexlab{a}})\citenamefont {Ferraro}, \citenamefont {Tak{\'a}cs}, \citenamefont {Piatti}, \citenamefont {Cavanna}, \citenamefont {Depalo}, \citenamefont {Aliotta}, \citenamefont {Bemmerer}, \citenamefont {Best}, \citenamefont {Boeltzig}, \citenamefont {Broggini} \emph {et~al.}}]{ferraro2018direct}%
  \BibitemOpen
  \bibfield  {author} {\bibinfo {author} {\bibfnamefont {F.}~\bibnamefont {Ferraro}}, \bibinfo {author} {\bibfnamefont {M.~P.}\ \bibnamefont {Tak{\'a}cs}}, \bibinfo {author} {\bibfnamefont {D.}~\bibnamefont {Piatti}}, \bibinfo {author} {\bibfnamefont {F.}~\bibnamefont {Cavanna}}, \bibinfo {author} {\bibfnamefont {R.}~\bibnamefont {Depalo}}, \bibinfo {author} {\bibfnamefont {M.}~\bibnamefont {Aliotta}}, \bibinfo {author} {\bibfnamefont {D.}~\bibnamefont {Bemmerer}}, \bibinfo {author} {\bibfnamefont {A.}~\bibnamefont {Best}}, \bibinfo {author} {\bibfnamefont {A.}~\bibnamefont {Boeltzig}}, \bibinfo {author} {\bibfnamefont {C.}~\bibnamefont {Broggini}}, \emph {et~al.},\ }\href {https://doi.org/10.1103/PhysRevLett.121.172701} {\bibfield  {journal} {\bibinfo  {journal} {Physical Review Letters}\ }\textbf {\bibinfo {volume} {121}},\ \bibinfo {pages} {172701} (\bibinfo {year} {2018}{\natexlab{a}})}\BibitemShut {NoStop}%
\bibitem [{\citenamefont {Williams}\ \emph {et~al.}(2020)\citenamefont {Williams}, \citenamefont {Lennarz}, \citenamefont {Laird}, \citenamefont {Battino}, \citenamefont {Jos{\'e}}, \citenamefont {Connolly}, \citenamefont {Ruiz}, \citenamefont {Chen}, \citenamefont {Davids}, \citenamefont {Esker} \emph {et~al.}}]{williams2020first}%
  \BibitemOpen
  \bibfield  {author} {\bibinfo {author} {\bibfnamefont {M.}~\bibnamefont {Williams}}, \bibinfo {author} {\bibfnamefont {A.}~\bibnamefont {Lennarz}}, \bibinfo {author} {\bibfnamefont {A.}~\bibnamefont {Laird}}, \bibinfo {author} {\bibfnamefont {U.}~\bibnamefont {Battino}}, \bibinfo {author} {\bibfnamefont {J.}~\bibnamefont {Jos{\'e}}}, \bibinfo {author} {\bibfnamefont {D.}~\bibnamefont {Connolly}}, \bibinfo {author} {\bibfnamefont {C.}~\bibnamefont {Ruiz}}, \bibinfo {author} {\bibfnamefont {A.}~\bibnamefont {Chen}}, \bibinfo {author} {\bibfnamefont {B.}~\bibnamefont {Davids}}, \bibinfo {author} {\bibfnamefont {N.}~\bibnamefont {Esker}}, \emph {et~al.},\ }\href {https://doi.org/10.1103/PhysRevC.102.035801} {\bibfield  {journal} {\bibinfo  {journal} {Physical Review C}\ }\textbf {\bibinfo {volume} {102}},\ \bibinfo {pages} {035801} (\bibinfo {year} {2020})}\BibitemShut {NoStop}%
\bibitem [{\citenamefont {Masha}\ \emph {et~al.}(2023)\citenamefont {Masha}, \citenamefont {Barbieri}, \citenamefont {Skowronski}, \citenamefont {Aliotta}, \citenamefont {Ananna}, \citenamefont {Barile}, \citenamefont {Bemmerer}, \citenamefont {Best}, \citenamefont {Boeltzig}, \citenamefont {Broggini}, \citenamefont {Bruno} \emph {et~al.}}]{masha2023prc}%
  \BibitemOpen
  \bibfield  {author} {\bibinfo {author} {\bibfnamefont {E.}~\bibnamefont {Masha}}, \bibinfo {author} {\bibfnamefont {L.}~\bibnamefont {Barbieri}}, \bibinfo {author} {\bibfnamefont {J.}~\bibnamefont {Skowronski}}, \bibinfo {author} {\bibfnamefont {M.}~\bibnamefont {Aliotta}}, \bibinfo {author} {\bibfnamefont {C.}~\bibnamefont {Ananna}}, \bibinfo {author} {\bibfnamefont {F.}~\bibnamefont {Barile}}, \bibinfo {author} {\bibfnamefont {D.}~\bibnamefont {Bemmerer}}, \bibinfo {author} {\bibfnamefont {A.}~\bibnamefont {Best}}, \bibinfo {author} {\bibfnamefont {A.}~\bibnamefont {Boeltzig}}, \bibinfo {author} {\bibfnamefont {C.}~\bibnamefont {Broggini}}, \bibinfo {author} {\bibfnamefont {C.~G.}\ \bibnamefont {Bruno}}, \emph {et~al.} (\bibinfo {collaboration} {LUNA collaboration}),\ }\href {https://doi.org/10.1103/PhysRevC.108.L052801} {\bibfield  {journal} {\bibinfo  {journal} {Physical Review C}\ }\textbf {\bibinfo {volume} {108}},\ \bibinfo {pages} {L052801} (\bibinfo {year} {2023})}\BibitemShut {NoStop}%
\bibitem [{\citenamefont {Tak\'acs}\ \emph {et~al.}(2024)\citenamefont {Tak\'acs}, \citenamefont {Ferraro}, \citenamefont {Piatti}, \citenamefont {Skowronski}, \citenamefont {Aliotta}, \citenamefont {Ananna}, \citenamefont {Barbieri}, \citenamefont {Barile}, \citenamefont {Bemmerer}, \citenamefont {Best} \emph {et~al.}}]{Takacs24-PRC}%
  \BibitemOpen
  \bibfield  {author} {\bibinfo {author} {\bibfnamefont {M.~P.}\ \bibnamefont {Tak\'acs}}, \bibinfo {author} {\bibfnamefont {F.}~\bibnamefont {Ferraro}}, \bibinfo {author} {\bibfnamefont {D.}~\bibnamefont {Piatti}}, \bibinfo {author} {\bibfnamefont {J.}~\bibnamefont {Skowronski}}, \bibinfo {author} {\bibfnamefont {M.}~\bibnamefont {Aliotta}}, \bibinfo {author} {\bibfnamefont {C.}~\bibnamefont {Ananna}}, \bibinfo {author} {\bibfnamefont {L.}~\bibnamefont {Barbieri}}, \bibinfo {author} {\bibfnamefont {F.}~\bibnamefont {Barile}}, \bibinfo {author} {\bibfnamefont {D.}~\bibnamefont {Bemmerer}}, \bibinfo {author} {\bibfnamefont {A.}~\bibnamefont {Best}}, \emph {et~al.} (\bibinfo {collaboration} {LUNA Collaboration}),\ }\href {https://doi.org/10.1103/PhysRevC.109.064627} {\bibfield  {journal} {\bibinfo  {journal} {Physical Review C}\ }\textbf {\bibinfo {volume} {109}},\ \bibinfo {pages} {064627} (\bibinfo {year} {2024})}\BibitemShut {NoStop}%
\bibitem [{\citenamefont {Weiss}\ and\ \citenamefont {Truran}(1990)}]{weiss199022}%
  \BibitemOpen
  \bibfield  {author} {\bibinfo {author} {\bibfnamefont {A.}~\bibnamefont {Weiss}}\ and\ \bibinfo {author} {\bibfnamefont {J.~W.}\ \bibnamefont {Truran}},\ }\href {https://adsabs.harvard.edu/full/1990A%26A...238..178W} {\bibfield  {journal} {\bibinfo  {journal} {Astronomy and Astrophysics}\ }\textbf {\bibinfo {volume} {238}},\ \bibinfo {pages} {178} (\bibinfo {year} {1990})}\BibitemShut {NoStop}%
\bibitem [{\citenamefont {Basunia}(2015)}]{BASUNIA201569}%
  \BibitemOpen
  \bibfield  {author} {\bibinfo {author} {\bibfnamefont {M.~S.}\ \bibnamefont {Basunia}},\ }\href {https://doi.org/https://doi.org/10.1016/j.nds.2015.07.002} {\bibfield  {journal} {\bibinfo  {journal} {Nuclear Data Sheets}\ }\textbf {\bibinfo {volume} {127}},\ \bibinfo {pages} {69} (\bibinfo {year} {2015})}\BibitemShut {NoStop}%
\bibitem [{\citenamefont {Foug\`eres}\ \emph {et~al.}(2023)\citenamefont {Foug\`eres} \emph {et~al.}}]{Fougeres:2022upo}%
  \BibitemOpen
  \bibfield  {author} {\bibinfo {author} {\bibfnamefont {C.}~\bibnamefont {Foug\`eres}} \emph {et~al.},\ }\href {https://doi.org/10.1038/s41467-023-40121-3} {\bibfield  {journal} {\bibinfo  {journal} {Nature Communications}\ }\textbf {\bibinfo {volume} {14}},\ \bibinfo {pages} {4536} (\bibinfo {year} {2023})}\BibitemShut {NoStop}%
\bibitem [{\citenamefont {Fryer}\ \emph {et~al.}(2019)\citenamefont {Fryer}, \citenamefont {Timmes}, \citenamefont {Hungerford}, \citenamefont {Couture}, \citenamefont {Adams}, \citenamefont {Aoki}, \citenamefont {Arcones}, \citenamefont {Arnett}, \citenamefont {Auchettl}, \citenamefont {Avila} \emph {et~al.}}]{fryer2019catching}%
  \BibitemOpen
  \bibfield  {author} {\bibinfo {author} {\bibfnamefont {C.~L.}\ \bibnamefont {Fryer}}, \bibinfo {author} {\bibfnamefont {F.}~\bibnamefont {Timmes}}, \bibinfo {author} {\bibfnamefont {A.~L.}\ \bibnamefont {Hungerford}}, \bibinfo {author} {\bibfnamefont {A.}~\bibnamefont {Couture}}, \bibinfo {author} {\bibfnamefont {F.}~\bibnamefont {Adams}}, \bibinfo {author} {\bibfnamefont {W.}~\bibnamefont {Aoki}}, \bibinfo {author} {\bibfnamefont {A.}~\bibnamefont {Arcones}}, \bibinfo {author} {\bibfnamefont {D.}~\bibnamefont {Arnett}}, \bibinfo {author} {\bibfnamefont {K.}~\bibnamefont {Auchettl}}, \bibinfo {author} {\bibfnamefont {M.}~\bibnamefont {Avila}}, \emph {et~al.},\ }\href {https://orbit.dtu.dk/en/publications/catching-element-formation-in-the-act-the-case-for-a-new-mev-gamm} {\bibfield  {journal} {\bibinfo  {journal} {Bulletin of the American Astronomical Society}\ }\textbf {\bibinfo {volume} {51}} (\bibinfo {year} {2019})}\BibitemShut {NoStop}%
\bibitem [{\citenamefont {Woosley}\ and\ \citenamefont {Janka}(2005)}]{woosley2005physics}%
  \BibitemOpen
  \bibfield  {author} {\bibinfo {author} {\bibfnamefont {S.}~\bibnamefont {Woosley}}\ and\ \bibinfo {author} {\bibfnamefont {T.}~\bibnamefont {Janka}},\ }\href {https://www.nature.com/articles/nphys172} {\bibfield  {journal} {\bibinfo  {journal} {Nature Physics}\ }\textbf {\bibinfo {volume} {1}},\ \bibinfo {pages} {147} (\bibinfo {year} {2005})}\BibitemShut {NoStop}%
\bibitem [{\citenamefont {Janka}\ \emph {et~al.}(2007)\citenamefont {Janka}, \citenamefont {Langanke}, \citenamefont {Marek}, \citenamefont {Mart{\'\i}nez-Pinedo},\ and\ \citenamefont {M{\"u}ller}}]{janka2007theory}%
  \BibitemOpen
  \bibfield  {author} {\bibinfo {author} {\bibfnamefont {H.-T.}\ \bibnamefont {Janka}}, \bibinfo {author} {\bibfnamefont {K.}~\bibnamefont {Langanke}}, \bibinfo {author} {\bibfnamefont {A.}~\bibnamefont {Marek}}, \bibinfo {author} {\bibfnamefont {G.}~\bibnamefont {Mart{\'\i}nez-Pinedo}},\ and\ \bibinfo {author} {\bibfnamefont {B.}~\bibnamefont {M{\"u}ller}},\ }\href {https://doi.org/10.1016/j.physrep.2007.02.002} {\bibfield  {journal} {\bibinfo  {journal} {Physics Reports}\ }\textbf {\bibinfo {volume} {442}},\ \bibinfo {pages} {38} (\bibinfo {year} {2007})}\BibitemShut {NoStop}%
\bibitem [{\citenamefont {Junde}\ \emph {et~al.}(2011)\citenamefont {Junde}, \citenamefont {Su},\ and\ \citenamefont {Dong}}]{JUNDE20111513}%
  \BibitemOpen
  \bibfield  {author} {\bibinfo {author} {\bibfnamefont {H.}~\bibnamefont {Junde}}, \bibinfo {author} {\bibfnamefont {H.}~\bibnamefont {Su}},\ and\ \bibinfo {author} {\bibfnamefont {Y.}~\bibnamefont {Dong}},\ }\href {https://doi.org/https://doi.org/10.1016/j.nds.2011.04.004} {\bibfield  {journal} {\bibinfo  {journal} {Nuclear Data Sheets}\ }\textbf {\bibinfo {volume} {112}},\ \bibinfo {pages} {1513} (\bibinfo {year} {2011})}\BibitemShut {NoStop}%
\bibitem [{\citenamefont {Amari}(2008)}]{amari2008sodium}%
  \BibitemOpen
  \bibfield  {author} {\bibinfo {author} {\bibfnamefont {S.}~\bibnamefont {Amari}},\ }\href {https://iopscience.iop.org/article/10.1088/0004-637X/690/2/1424/meta} {\bibfield  {journal} {\bibinfo  {journal} {The Astrophysical Journal}\ }\textbf {\bibinfo {volume} {690}},\ \bibinfo {pages} {1424} (\bibinfo {year} {2008})}\BibitemShut {NoStop}%
\bibitem [{\citenamefont {Amari}\ \emph {et~al.}(1990)\citenamefont {Amari}, \citenamefont {Anders}, \citenamefont {Virag},\ and\ \citenamefont {Zinner}}]{amari1990interstellar}%
  \BibitemOpen
  \bibfield  {author} {\bibinfo {author} {\bibfnamefont {S.}~\bibnamefont {Amari}}, \bibinfo {author} {\bibfnamefont {E.}~\bibnamefont {Anders}}, \bibinfo {author} {\bibfnamefont {A.}~\bibnamefont {Virag}},\ and\ \bibinfo {author} {\bibfnamefont {E.}~\bibnamefont {Zinner}},\ }\href {https://www.nature.com/articles/345238a0} {\bibfield  {journal} {\bibinfo  {journal} {Nature}\ }\textbf {\bibinfo {volume} {345}},\ \bibinfo {pages} {238} (\bibinfo {year} {1990})}\BibitemShut {NoStop}%
\bibitem [{\citenamefont {Amari}\ \emph {et~al.}(2001)\citenamefont {Amari}, \citenamefont {Nittler}, \citenamefont {Zinner}, \citenamefont {Gallino}, \citenamefont {Lugaro},\ and\ \citenamefont {Lewis}}]{Amari_2001}%
  \BibitemOpen
  \bibfield  {author} {\bibinfo {author} {\bibfnamefont {S.}~\bibnamefont {Amari}}, \bibinfo {author} {\bibfnamefont {L.~R.}\ \bibnamefont {Nittler}}, \bibinfo {author} {\bibfnamefont {E.}~\bibnamefont {Zinner}}, \bibinfo {author} {\bibfnamefont {R.}~\bibnamefont {Gallino}}, \bibinfo {author} {\bibfnamefont {M.}~\bibnamefont {Lugaro}},\ and\ \bibinfo {author} {\bibfnamefont {R.~S.}\ \bibnamefont {Lewis}},\ }\href {https://doi.org/10.1086/318230} {\bibfield  {journal} {\bibinfo  {journal} {The Astrophysical Journal}\ }\textbf {\bibinfo {volume} {546}},\ \bibinfo {pages} {248} (\bibinfo {year} {2001})}\BibitemShut {NoStop}%
\bibitem [{\citenamefont {G{\"o}rres}\ \emph {et~al.}(1982)\citenamefont {G{\"o}rres}, \citenamefont {Rolfs}, \citenamefont {Schmalbrock}, \citenamefont {Trautvetter},\ and\ \citenamefont {Keinonen}}]{gorres1982search}%
  \BibitemOpen
  \bibfield  {author} {\bibinfo {author} {\bibfnamefont {J.}~\bibnamefont {G{\"o}rres}}, \bibinfo {author} {\bibfnamefont {C.}~\bibnamefont {Rolfs}}, \bibinfo {author} {\bibfnamefont {P.}~\bibnamefont {Schmalbrock}}, \bibinfo {author} {\bibfnamefont {H.}~\bibnamefont {Trautvetter}},\ and\ \bibinfo {author} {\bibfnamefont {J.}~\bibnamefont {Keinonen}},\ }\href {https://www.sciencedirect.com/science/article/pii/0375947482904894} {\bibfield  {journal} {\bibinfo  {journal} {Nuclear Physics A}\ }\textbf {\bibinfo {volume} {385}},\ \bibinfo {pages} {57} (\bibinfo {year} {1982})}\BibitemShut {NoStop}%
\bibitem [{\citenamefont {Berg}\ \emph {et~al.}(1977)\citenamefont {Berg}, \citenamefont {Hietzke}, \citenamefont {Rolfs},\ and\ \citenamefont {Winkler}}]{berg1977proton}%
  \BibitemOpen
  \bibfield  {author} {\bibinfo {author} {\bibfnamefont {H.}~\bibnamefont {Berg}}, \bibinfo {author} {\bibfnamefont {W.}~\bibnamefont {Hietzke}}, \bibinfo {author} {\bibfnamefont {C.}~\bibnamefont {Rolfs}},\ and\ \bibinfo {author} {\bibfnamefont {H.}~\bibnamefont {Winkler}},\ }\href {https://www.sciencedirect.com/science/article/pii/0375947477901658} {\bibfield  {journal} {\bibinfo  {journal} {Nuclear Physics A}\ }\textbf {\bibinfo {volume} {276}},\ \bibinfo {pages} {168} (\bibinfo {year} {1977})}\BibitemShut {NoStop}%
\bibitem [{\citenamefont {G{\"o}rres}\ \emph {et~al.}(1983)\citenamefont {G{\"o}rres}, \citenamefont {Becker}, \citenamefont {Buchmann}, \citenamefont {Rolfs}, \citenamefont {Schmalbrock}, \citenamefont {Trautvetter}, \citenamefont {Vlieks}, \citenamefont {Hammer},\ and\ \citenamefont {Donoghue}}]{gorres1983proton}%
  \BibitemOpen
  \bibfield  {author} {\bibinfo {author} {\bibfnamefont {J.}~\bibnamefont {G{\"o}rres}}, \bibinfo {author} {\bibfnamefont {H.}~\bibnamefont {Becker}}, \bibinfo {author} {\bibfnamefont {L.}~\bibnamefont {Buchmann}}, \bibinfo {author} {\bibfnamefont {C.}~\bibnamefont {Rolfs}}, \bibinfo {author} {\bibfnamefont {P.}~\bibnamefont {Schmalbrock}}, \bibinfo {author} {\bibfnamefont {H.}~\bibnamefont {Trautvetter}}, \bibinfo {author} {\bibfnamefont {A.}~\bibnamefont {Vlieks}}, \bibinfo {author} {\bibfnamefont {J.}~\bibnamefont {Hammer}},\ and\ \bibinfo {author} {\bibfnamefont {T.}~\bibnamefont {Donoghue}},\ }\href {https://www.sciencedirect.com/science/article/pii/0375947483905882} {\bibfield  {journal} {\bibinfo  {journal} {Nuclear Physics A}\ }\textbf {\bibinfo {volume} {408}},\ \bibinfo {pages} {372} (\bibinfo {year} {1983})}\BibitemShut {NoStop}%
\bibitem [{\citenamefont {Becker}\ \emph {et~al.}(1992)\citenamefont {Becker}, \citenamefont {Ebbing}, \citenamefont {Schulte}, \citenamefont {W{\"u}stenbecker}, \citenamefont {Berheide}, \citenamefont {Buschmann}, \citenamefont {Rolfs}, \citenamefont {Mitchell},\ and\ \citenamefont {Schweitzer}}]{becker1992low}%
  \BibitemOpen
  \bibfield  {author} {\bibinfo {author} {\bibfnamefont {H.}~\bibnamefont {Becker}}, \bibinfo {author} {\bibfnamefont {H.}~\bibnamefont {Ebbing}}, \bibinfo {author} {\bibfnamefont {W.}~\bibnamefont {Schulte}}, \bibinfo {author} {\bibfnamefont {S.}~\bibnamefont {W{\"u}stenbecker}}, \bibinfo {author} {\bibfnamefont {M.}~\bibnamefont {Berheide}}, \bibinfo {author} {\bibfnamefont {M.}~\bibnamefont {Buschmann}}, \bibinfo {author} {\bibfnamefont {C.}~\bibnamefont {Rolfs}}, \bibinfo {author} {\bibfnamefont {G.}~\bibnamefont {Mitchell}},\ and\ \bibinfo {author} {\bibfnamefont {J.}~\bibnamefont {Schweitzer}},\ }\href {https://link.springer.com/article/10.1007/BF01291537} {\bibfield  {journal} {\bibinfo  {journal} {Zeitschrift f{\"u}r Physik A Hadrons and Nuclei}\ }\textbf {\bibinfo {volume} {343}},\ \bibinfo {pages} {361} (\bibinfo {year} {1992})}\BibitemShut {NoStop}%
\bibitem [{\citenamefont {Aliotta}\ \emph {et~al.}(2022)\citenamefont {Aliotta}, \citenamefont {Boeltzig}, \citenamefont {Depalo},\ and\ \citenamefont {Gy{\"u}rky}}]{aliotta2022exploring}%
  \BibitemOpen
  \bibfield  {author} {\bibinfo {author} {\bibfnamefont {M.}~\bibnamefont {Aliotta}}, \bibinfo {author} {\bibfnamefont {A.}~\bibnamefont {Boeltzig}}, \bibinfo {author} {\bibfnamefont {R.}~\bibnamefont {Depalo}},\ and\ \bibinfo {author} {\bibfnamefont {G.}~\bibnamefont {Gy{\"u}rky}},\ }\href {https://www.annualreviews.org/content/journals/10.1146/annurev-nucl-110221-103625} {\bibfield  {journal} {\bibinfo  {journal} {Annual Review of Nuclear and Particle Science}\ }\textbf {\bibinfo {volume} {72}},\ \bibinfo {pages} {177} (\bibinfo {year} {2022})}\BibitemShut {NoStop}%
\bibitem [{\citenamefont {{Caciolli}}\ \emph {et~al.}(2009)\citenamefont {{Caciolli}}, \citenamefont {{Agostino}}, \citenamefont {{Bemmerer}}, \citenamefont {{Bonetti}}, \citenamefont {{Broggini}}, \citenamefont {{Confortola}}, \citenamefont {{Corvisiero}}, \citenamefont {{Costantini}}, \citenamefont {{Elekes}}, \citenamefont {{Formicola}} \emph {et~al.}}]{Caciolli09-EPJA}%
  \BibitemOpen
  \bibfield  {author} {\bibinfo {author} {\bibfnamefont {A.}~\bibnamefont {{Caciolli}}}, \bibinfo {author} {\bibfnamefont {L.}~\bibnamefont {{Agostino}}}, \bibinfo {author} {\bibfnamefont {D.}~\bibnamefont {{Bemmerer}}}, \bibinfo {author} {\bibfnamefont {R.}~\bibnamefont {{Bonetti}}}, \bibinfo {author} {\bibfnamefont {C.}~\bibnamefont {{Broggini}}}, \bibinfo {author} {\bibfnamefont {F.}~\bibnamefont {{Confortola}}}, \bibinfo {author} {\bibfnamefont {P.}~\bibnamefont {{Corvisiero}}}, \bibinfo {author} {\bibfnamefont {H.}~\bibnamefont {{Costantini}}}, \bibinfo {author} {\bibfnamefont {Z.}~\bibnamefont {{Elekes}}}, \bibinfo {author} {\bibfnamefont {A.}~\bibnamefont {{Formicola}}}, \emph {et~al.},\ }\href {https://doi.org/10.1140/epja/i2008-10706-3} {\bibfield  {journal} {\bibinfo  {journal} {The European Physical Journal A}\ }\textbf {\bibinfo {volume} {39}},\ \bibinfo {pages} {179} (\bibinfo {year} {2009})}\BibitemShut {NoStop}%
\bibitem [{\citenamefont {{Sz{\"u}cs}}\ \emph {et~al.}(2010)\citenamefont {{Sz{\"u}cs}}, \citenamefont {{Bemmerer}}, \citenamefont {{Broggini}}, \citenamefont {{Caciolli}}, \citenamefont {{Confortola}}, \citenamefont {{Corvisiero}}, \citenamefont {{Elekes}}, \citenamefont {{Formicola}}, \citenamefont {{F{\"u}l{\"o}p}}, \citenamefont {{Gervino}} \emph {et~al.}}]{Szucs10-EPJA}%
  \BibitemOpen
  \bibfield  {author} {\bibinfo {author} {\bibfnamefont {T.}~\bibnamefont {{Sz{\"u}cs}}}, \bibinfo {author} {\bibfnamefont {D.}~\bibnamefont {{Bemmerer}}}, \bibinfo {author} {\bibfnamefont {C.}~\bibnamefont {{Broggini}}}, \bibinfo {author} {\bibfnamefont {A.}~\bibnamefont {{Caciolli}}}, \bibinfo {author} {\bibfnamefont {F.}~\bibnamefont {{Confortola}}}, \bibinfo {author} {\bibfnamefont {P.}~\bibnamefont {{Corvisiero}}}, \bibinfo {author} {\bibfnamefont {Z.}~\bibnamefont {{Elekes}}}, \bibinfo {author} {\bibfnamefont {A.}~\bibnamefont {{Formicola}}}, \bibinfo {author} {\bibfnamefont {Z.}~\bibnamefont {{F{\"u}l{\"o}p}}}, \bibinfo {author} {\bibfnamefont {G.}~\bibnamefont {{Gervino}}}, \emph {et~al.},\ }\href {https://doi.org/10.1140/epja/i2010-10967-1} {\bibfield  {journal} {\bibinfo  {journal} {The European Physical Journal A}\ }\textbf {\bibinfo {volume} {44}},\ \bibinfo {pages} {513} (\bibinfo {year} {2010})}\BibitemShut {NoStop}%
\bibitem [{\citenamefont {Cavanna}\ \emph {et~al.}(2014)\citenamefont {Cavanna}, \citenamefont {Depalo}, \citenamefont {Menzel}, \citenamefont {Aliotta}, \citenamefont {Anders}, \citenamefont {Bemmerer}, \citenamefont {Broggini}, \citenamefont {Bruno}, \citenamefont {Caciolli} \emph {et~al.}}]{cavanna2014epja}%
  \BibitemOpen
  \bibfield  {author} {\bibinfo {author} {\bibfnamefont {F.}~\bibnamefont {Cavanna}}, \bibinfo {author} {\bibfnamefont {R.}~\bibnamefont {Depalo}}, \bibinfo {author} {\bibfnamefont {M.~L.}\ \bibnamefont {Menzel}}, \bibinfo {author} {\bibfnamefont {M.}~\bibnamefont {Aliotta}}, \bibinfo {author} {\bibfnamefont {M.}~\bibnamefont {Anders}}, \bibinfo {author} {\bibfnamefont {D.}~\bibnamefont {Bemmerer}}, \bibinfo {author} {\bibfnamefont {C.}~\bibnamefont {Broggini}}, \bibinfo {author} {\bibfnamefont {C.}~\bibnamefont {Bruno}}, \bibinfo {author} {\bibfnamefont {A.}~\bibnamefont {Caciolli}}, \emph {et~al.},\ }\href {https://link.springer.com/article/10.1140/epja/i2014-14179-5} {\bibfield  {journal} {\bibinfo  {journal} {The European Physical Journal A}\ }\textbf {\bibinfo {volume} {50}},\ \bibinfo {pages} {1} (\bibinfo {year} {2014})}\BibitemShut {NoStop}%
\bibitem [{\citenamefont {Ferraro}\ \emph {et~al.}(2018{\natexlab{b}})\citenamefont {Ferraro}, \citenamefont {Tak{\'a}cs}, \citenamefont {Piatti}, \citenamefont {Mossa}, \citenamefont {Aliotta}, \citenamefont {Bemmerer}, \citenamefont {Best}, \citenamefont {Boeltzig}, \citenamefont {Broggini} \emph {et~al.}}]{ferraro2018epja}%
  \BibitemOpen
  \bibfield  {author} {\bibinfo {author} {\bibfnamefont {F.}~\bibnamefont {Ferraro}}, \bibinfo {author} {\bibfnamefont {M.}~\bibnamefont {Tak{\'a}cs}}, \bibinfo {author} {\bibfnamefont {D.}~\bibnamefont {Piatti}}, \bibinfo {author} {\bibfnamefont {V.}~\bibnamefont {Mossa}}, \bibinfo {author} {\bibfnamefont {M.}~\bibnamefont {Aliotta}}, \bibinfo {author} {\bibfnamefont {D.}~\bibnamefont {Bemmerer}}, \bibinfo {author} {\bibfnamefont {A.}~\bibnamefont {Best}}, \bibinfo {author} {\bibfnamefont {A.}~\bibnamefont {Boeltzig}}, \bibinfo {author} {\bibfnamefont {C.}~\bibnamefont {Broggini}}, \emph {et~al.},\ }\href {https://link.springer.com/article/10.1140/epja/i2018-12476-7} {\bibfield  {journal} {\bibinfo  {journal} {The European Physical Journal A}\ }\textbf {\bibinfo {volume} {54}},\ \bibinfo {pages} {1} (\bibinfo {year} {2018}{\natexlab{b}})}\BibitemShut {NoStop}%
\bibitem [{\citenamefont {Daigle}\ \emph {et~al.}(2016)\citenamefont {Daigle}, \citenamefont {Kelly}, \citenamefont {Champagne}, \citenamefont {Buckner}, \citenamefont {Iliadis},\ and\ \citenamefont {Howard}}]{daigle2016}%
  \BibitemOpen
  \bibfield  {author} {\bibinfo {author} {\bibfnamefont {S.}~\bibnamefont {Daigle}}, \bibinfo {author} {\bibfnamefont {K.~J.}\ \bibnamefont {Kelly}}, \bibinfo {author} {\bibfnamefont {A.~E.}\ \bibnamefont {Champagne}}, \bibinfo {author} {\bibfnamefont {M.~Q.}\ \bibnamefont {Buckner}}, \bibinfo {author} {\bibfnamefont {C.}~\bibnamefont {Iliadis}},\ and\ \bibinfo {author} {\bibfnamefont {C.}~\bibnamefont {Howard}},\ }\href {https://doi.org/10.1103/PhysRevC.94.025803} {\bibfield  {journal} {\bibinfo  {journal} {Physical Review C}\ }\textbf {\bibinfo {volume} {94}},\ \bibinfo {pages} {025803} (\bibinfo {year} {2016})}\BibitemShut {NoStop}%
\bibitem [{\citenamefont {Masha}\ \emph {et~al.}(2025)\citenamefont {Masha}, \citenamefont {Casaburo}, \citenamefont {Sidhu}, \citenamefont {Aliotta}, \citenamefont {Ananna}, \citenamefont {Barbieri}, \citenamefont {Barile}, \citenamefont {Bemmerer}, \citenamefont {Best}, \citenamefont {Boeltzig} \emph {et~al.}}]{masha2024epja}%
  \BibitemOpen
  \bibfield  {author} {\bibinfo {author} {\bibfnamefont {E.}~\bibnamefont {Masha}}, \bibinfo {author} {\bibfnamefont {F.}~\bibnamefont {Casaburo}}, \bibinfo {author} {\bibfnamefont {R.~S.}\ \bibnamefont {Sidhu}}, \bibinfo {author} {\bibfnamefont {M.}~\bibnamefont {Aliotta}}, \bibinfo {author} {\bibfnamefont {C.}~\bibnamefont {Ananna}}, \bibinfo {author} {\bibfnamefont {L.}~\bibnamefont {Barbieri}}, \bibinfo {author} {\bibfnamefont {F.}~\bibnamefont {Barile}}, \bibinfo {author} {\bibfnamefont {D.}~\bibnamefont {Bemmerer}}, \bibinfo {author} {\bibfnamefont {A.}~\bibnamefont {Best}}, \bibinfo {author} {\bibfnamefont {A.}~\bibnamefont {Boeltzig}}, \emph {et~al.},\ }\href {https://doi.org/10.1140/epja/s10050-025-01512-w} {\bibfield  {journal} {\bibinfo  {journal} {The European Physical Journal A}\ }\textbf {\bibinfo {volume} {61}},\ \bibinfo {pages} {45} (\bibinfo {year} {2025})}\BibitemShut {NoStop}%
\bibitem [{\citenamefont {Sidhu}\ \emph {et~al.}(2024)\citenamefont {Sidhu}, \citenamefont {Casaburo}, \citenamefont {Masha} \emph {et~al.}}]{Sidhu_supplement}%
  \BibitemOpen
  \bibfield  {author} {\bibinfo {author} {\bibfnamefont {R.~S.}\ \bibnamefont {Sidhu}}, \bibinfo {author} {\bibfnamefont {F.}~\bibnamefont {Casaburo}}, \bibinfo {author} {\bibfnamefont {E.}~\bibnamefont {Masha}}, \emph {et~al.},\ }\href@noop {} {} (\bibinfo {year} {2024}),\ \bibinfo {note} {{Supplemental Material for more details}}\BibitemShut {NoStop}%
\bibitem [{\citenamefont {Fowler}\ \emph {et~al.}(1948)\citenamefont {Fowler}, \citenamefont {Lauritsen},\ and\ \citenamefont {Lauritsen}}]{RevModPhys.20.236}%
  \BibitemOpen
  \bibfield  {author} {\bibinfo {author} {\bibfnamefont {W.~A.}\ \bibnamefont {Fowler}}, \bibinfo {author} {\bibfnamefont {C.~C.}\ \bibnamefont {Lauritsen}},\ and\ \bibinfo {author} {\bibfnamefont {T.}~\bibnamefont {Lauritsen}},\ }\href {https://doi.org/10.1103/RevModPhys.20.236} {\bibfield  {journal} {\bibinfo  {journal} {Review of Modern Physics}\ }\textbf {\bibinfo {volume} {20}},\ \bibinfo {pages} {236} (\bibinfo {year} {1948})}\BibitemShut {NoStop}%
\bibitem [{\citenamefont {Rolfs}\ and\ \citenamefont {Rodney}(1988)}]{rolfs1988cauldrons}%
  \BibitemOpen
  \bibfield  {author} {\bibinfo {author} {\bibfnamefont {C.~E.}\ \bibnamefont {Rolfs}}\ and\ \bibinfo {author} {\bibfnamefont {W.~S.}\ \bibnamefont {Rodney}},\ }\href {https://press.uchicago.edu/ucp/books/book/chicago/C/bo5964636.html} {\emph {\bibinfo {title} {{Cauldrons in the cosmos: Nuclear astrophysics}}}}\ (\bibinfo  {publisher} {University of Chicago press},\ \bibinfo {year} {1988})\BibitemShut {NoStop}%
\bibitem [{\citenamefont {Bemmerer}\ \emph {et~al.}(2018)\citenamefont {Bemmerer}, \citenamefont {Cavanna}, \citenamefont {Depalo}, \citenamefont {Aliotta}, \citenamefont {Anders}, \citenamefont {Boeltzig}, \citenamefont {Broggini}, \citenamefont {Bruno}, \citenamefont {Caciolli}, \citenamefont {Chillery} \emph {et~al.}}]{bemmerer2018effect}%
  \BibitemOpen
  \bibfield  {author} {\bibinfo {author} {\bibfnamefont {D.}~\bibnamefont {Bemmerer}}, \bibinfo {author} {\bibfnamefont {F.}~\bibnamefont {Cavanna}}, \bibinfo {author} {\bibfnamefont {R.}~\bibnamefont {Depalo}}, \bibinfo {author} {\bibfnamefont {M.}~\bibnamefont {Aliotta}}, \bibinfo {author} {\bibfnamefont {M.}~\bibnamefont {Anders}}, \bibinfo {author} {\bibfnamefont {A.}~\bibnamefont {Boeltzig}}, \bibinfo {author} {\bibfnamefont {C.}~\bibnamefont {Broggini}}, \bibinfo {author} {\bibfnamefont {C.}~\bibnamefont {Bruno}}, \bibinfo {author} {\bibfnamefont {A.}~\bibnamefont {Caciolli}}, \bibinfo {author} {\bibfnamefont {T.}~\bibnamefont {Chillery}}, \emph {et~al.},\ }\href {https://iopscience.iop.org/article/10.1209/0295-5075/122/52001} {\bibfield  {journal} {\bibinfo  {journal} {Europhysics Letters}\ }\textbf {\bibinfo {volume} {122}},\ \bibinfo {pages} {52001} (\bibinfo {year} {2018})}\BibitemShut {NoStop}%
\bibitem [{\citenamefont {Assenbaum}\ \emph {et~al.}(1987)\citenamefont {Assenbaum}, \citenamefont {Langanke},\ and\ \citenamefont {Rolfs}}]{assenbaum1987effects}%
  \BibitemOpen
  \bibfield  {author} {\bibinfo {author} {\bibfnamefont {H.}~\bibnamefont {Assenbaum}}, \bibinfo {author} {\bibfnamefont {K.}~\bibnamefont {Langanke}},\ and\ \bibinfo {author} {\bibfnamefont {C.}~\bibnamefont {Rolfs}},\ }\href {https://link.springer.com/article/10.1007/BF01289572} {\bibfield  {journal} {\bibinfo  {journal} {Zeitschrift f{\"u}r Physik A Atomic Nuclei}\ }\textbf {\bibinfo {volume} {327}},\ \bibinfo {pages} {461} (\bibinfo {year} {1987})}\BibitemShut {NoStop}%
\bibitem [{\citenamefont {Iliadis}(2023)}]{iliadis2023laboratory}%
  \BibitemOpen
  \bibfield  {author} {\bibinfo {author} {\bibfnamefont {C.}~\bibnamefont {Iliadis}},\ }\href {https://journals.aps.org/prc/abstract/10.1103/PhysRevC.107.044610} {\bibfield  {journal} {\bibinfo  {journal} {Physical Review C}\ }\textbf {\bibinfo {volume} {107}},\ \bibinfo {pages} {044610} (\bibinfo {year} {2023})}\BibitemShut {NoStop}%
\bibitem [{\citenamefont {Sallaska}\ \emph {et~al.}(2013)\citenamefont {Sallaska}, \citenamefont {Iliadis}, \citenamefont {Champange}, \citenamefont {Goriely}, \citenamefont {Starrfield},\ and\ \citenamefont {Timmes}}]{sallaska2013starlib}%
  \BibitemOpen
  \bibfield  {author} {\bibinfo {author} {\bibfnamefont {A.~L.}\ \bibnamefont {Sallaska}}, \bibinfo {author} {\bibfnamefont {C.}~\bibnamefont {Iliadis}}, \bibinfo {author} {\bibfnamefont {A.}~\bibnamefont {Champange}}, \bibinfo {author} {\bibfnamefont {S.}~\bibnamefont {Goriely}}, \bibinfo {author} {\bibfnamefont {S.}~\bibnamefont {Starrfield}},\ and\ \bibinfo {author} {\bibfnamefont {F.}~\bibnamefont {Timmes}},\ }\href {https://iopscience.iop.org/article/10.1088/0067-0049/207/1/18/meta} {\bibfield  {journal} {\bibinfo  {journal} {The Astrophysical Journal Supplement Series}\ }\textbf {\bibinfo {volume} {207}},\ \bibinfo {pages} {18} (\bibinfo {year} {2013})}\BibitemShut {NoStop}%
\bibitem [{\citenamefont {Iliadis}\ \emph {et~al.}(2010{\natexlab{a}})\citenamefont {Iliadis}, \citenamefont {Longland}, \citenamefont {Champagne},\ and\ \citenamefont {Coc}}]{iliadis2010charged}%
  \BibitemOpen
  \bibfield  {author} {\bibinfo {author} {\bibfnamefont {C.}~\bibnamefont {Iliadis}}, \bibinfo {author} {\bibfnamefont {R.}~\bibnamefont {Longland}}, \bibinfo {author} {\bibfnamefont {A.}~\bibnamefont {Champagne}},\ and\ \bibinfo {author} {\bibfnamefont {A.}~\bibnamefont {Coc}},\ }\href {https://www.sciencedirect.com/science/article/pii/S0375947410004203} {\bibfield  {journal} {\bibinfo  {journal} {Nuclear Physics A}\ }\textbf {\bibinfo {volume} {841}},\ \bibinfo {pages} {251} (\bibinfo {year} {2010}{\natexlab{a}})}\BibitemShut {NoStop}%
\bibitem [{\citenamefont {Iliadis}\ \emph {et~al.}(2010{\natexlab{b}})\citenamefont {Iliadis}, \citenamefont {Longland}, \citenamefont {Champagne}, \citenamefont {Coc},\ and\ \citenamefont {Fitzgerald}}]{ILIADIS2010_II}%
  \BibitemOpen
  \bibfield  {author} {\bibinfo {author} {\bibfnamefont {C.}~\bibnamefont {Iliadis}}, \bibinfo {author} {\bibfnamefont {R.}~\bibnamefont {Longland}}, \bibinfo {author} {\bibfnamefont {A.}~\bibnamefont {Champagne}}, \bibinfo {author} {\bibfnamefont {A.}~\bibnamefont {Coc}},\ and\ \bibinfo {author} {\bibfnamefont {R.}~\bibnamefont {Fitzgerald}},\ }\href {https://doi.org/https://doi.org/10.1016/j.nuclphysa.2010.04.009} {\bibfield  {journal} {\bibinfo  {journal} {Nuclear Physics A}\ }\textbf {\bibinfo {volume} {841}},\ \bibinfo {pages} {31} (\bibinfo {year} {2010}{\natexlab{b}})}\BibitemShut {NoStop}%
\bibitem [{\citenamefont {José}\ and\ \citenamefont {Hernanz}(1998)}]{Jose_1998}%
  \BibitemOpen
  \bibfield  {author} {\bibinfo {author} {\bibfnamefont {J.}~\bibnamefont {José}}\ and\ \bibinfo {author} {\bibfnamefont {M.}~\bibnamefont {Hernanz}},\ }\href {https://doi.org/10.1086/305244} {\bibfield  {journal} {\bibinfo  {journal} {The Astrophysical Journal}\ }\textbf {\bibinfo {volume} {494}},\ \bibinfo {pages} {680} (\bibinfo {year} {1998})}\BibitemShut {NoStop}%
\bibitem [{\citenamefont {José}(2016)}]{Jose16-Book}%
  \BibitemOpen
  \bibfield  {author} {\bibinfo {author} {\bibfnamefont {J.}~\bibnamefont {José}},\ }\href {https://www.taylorfrancis.com/books/mono/10.1201/b19165/stellar-explosions-jordi-jose} {\emph {\bibinfo {title} {Stellar Explosions: Hydrodynamics and Nucleosynthesis}}}\ (\bibinfo  {publisher} {Taylor \& Francis},\ \bibinfo {address} {Boca Raton, Florida},\ \bibinfo {year} {2016})\BibitemShut {NoStop}%
\end{thebibliography}%

%\bibliography{apssamp}% Produces the bibliography via BibTeX.

\end{document}